\documentstyle[amsfonts,12pt,a4]{article}

\begin{document}

\begin{flushright}
quant-ph/9611019
\end{flushright}

\begin{flushleft}
Modern Physics Letters A, Vol.11, No. 26 (1996) 2095-2104\\[2.5cm]
\end{flushleft}

\begin{center}
\large
{\bf TIME-DEPENDENT PARASUPERSYMMETRY IN QUANTUM MECHANICS}
\normalsize
\\[10mm]
Boris F. Samsonov

Tomsk State University, 36 Lenin Ave. 634050, Tomsk, Russia\\
E-mail: samsonov@phys.tsu.tomsk.su\\[10mm]
\end{center}

\begin{abstract}
Parasupersymmetry of the one-dimensional time-dependent Schr\"o\-din\-ger
equ\-ation is established. It is intimately connected with a chain of the
time-dependent Darboux transformations. As an example a parasupersymmetric
model of nonrelativistic free particle with threefold degenerate discrete
spectrum of an integral of motion is constructed.
\end{abstract}

{\bf 1.} Supersymmetric quantum mechanics originally introduced by Witten 
\cite{Wit} for investigation of the supersymmetry breaking in quantum field
theories attracts now a considerable attention from the different points of
view (see a recent survey \cite{Coop}). Different generalizations of the
initial constructions are known. We can cite $N$-extended supersymmetric
quantum mechanics \cite{Pash}, higher-derivative supersymmetry in quantum
mechanics \cite{A}, and parasupersymmetric quantum mechanics \cite{pssqm}-%
\cite{BD}. It is worth stressing that all these constructions deal with a
stationary Schr\"o\-dinger equation and consequently can be referred to the
stationary supersymmetric quantum mechanics. The nonstationary
supersymmetric quantum mechanics needs to be developed. First steps in this
direction have been made in Refs. \cite{fiz} - \cite{BScs}.

An essential ingredient of the stationary supersymmetric quantum mechanics
constitutes the well-known Darboux transformation \cite{Darb} of the
stationary Schr\"o\-din\-ger equation in mathematics (see Ref. \cite{I}).
Recently the time-dependent Darboux transformation has been proposed \cite
{fiz} and on its base the supersymmetry of the nonstationary
Schr\"o\-din\-ger equation has been established \cite{BS}. This
transformation seems to be very fruitful for construction of the
nondispersive wave packets (coherent states) for anharmonic oscillator
Hamiltonians with equidistant and quasiequidistant spectra \cite{BScs}.

In this paper we establish the parasupersymmetry of the time-dependent
Schr\"o\-din\-ger equation which is intimately connected with a chain of the
time dependent Darboux transformations. With the help of these
transformations we can construct a chain of the new exactly solvable
nonstationary potentials starting from the initial potential even if it does
not depend on time. We introduce the notion of {\it completely reducible
chain of the Darboux transformations,} which is also valid for the
stationary case. We associate with this chain a nonlinear algebra of the
parasupersymmetric structure. As an example the application of two
consecutive time-dependent Darboux transformations to the free particle
Schr\"o\-din\-ger equation (zero potential) is made. This leads us to two
exactly solvable nonstationary potentials. For each potential the discrete
spectrum basis of the Hilbert space defined over solutions of the
nonstationary Schr\"o\-din\-ger equation is obtained. Three nilpotent
conserved supercharges together with the even symmetry operator formed a
nonlinear algebra of the parasupersymmetric structure are explicitly
constructed.

{\bf 2.} In this section we summarize following Refs. \cite{fiz}-\cite{BScs}
the main properties of the time-dependent Darboux transformation and
introduce the supersymmetry of the nonstationary Schr\"o\-din\-ger equation.

Let us consider two time-dependent Schr\"odinger equations 
\begin{equation}
\label{sc1}\left( i\partial _t-h_0\right) \psi \left( x,t\right) =0,\quad
h_0=-\partial _x^2+V_0\left( x,t\right) ,
\end{equation}
\begin{equation}
\label{sc2}\left( i\partial _t-h_1\right) \varphi \left( x,t\right) =0,\quad
h_1=-\partial _x^2+V_1\left( x,t\right) ,
\end{equation}
$$
\partial _x^2=\partial _x\cdot \partial _x,\quad \partial _x=\partial
/\partial x,\quad t\in {\Bbb R},\quad x\in R. 
$$
The interval $R=[a,b]$ for variable $x$ can be either finite or infinite.
Let $T_0$ and $T_1$ be linear spaces of the solutions of Eqs. (\ref{sc1})
and (\ref{sc2}) respectively defined over complex number field ${\Bbb C}$.
We assume that the solutions of Eq. (\ref{sc1}), called {\it the} {\it %
initial Schr\"o\-din\-ger equation,} are known. The problem the Darboux
transformation solves consists in derivation of such a set of potentials $%
V_1\left( x,t\right) $ that solutions $\varphi \left( x,t\right) $ of Eq. (%
\ref{sc2}) could be obtained by means of acting of a {\em first}-order
differential operator $L$ called {\it Darboux transformation operator} on
the solutions $\psi \left( x,t\right) $: $\varphi (x,t)=L\psi (x,t)$. Since
differential operators may have nontrivial kernels, we have the embedding $%
\widetilde{T}_1=\left\{ \varphi :\varphi =L\psi ,\forall
\psi \in T_0\right\} $ $\subset T_1$. This signifies that Eq. (\ref{sc2})
may have solutions which cannot be obtained by acting of operator $L$ on any
function $\psi \in T_0$. If necessary these solutions have to be obtained by
other means. The condition that the operator $L$ be first-order differential
operator is very restrictive for the potential $V_1(x,t)$. We have to look
for this potential together with the operator $L$. In order to achieve this
goal (i.e. to find the unknown functional coefficients of the operator $L$
and the function $V_1(x,t)$) we need to solve the operator equation \cite
{BScs} 
$$
L\left( i\partial _t-h_0\right) =\left( i\partial _t-h_1\right) L 
$$
in the space $T_0$. By virtue of Eq. (\ref{sc1}) the derivative of the
function $\psi (x,t)\in T_0$ with respect to $t$ can be replaced by $\left[
-\partial _x^2+V_0(x,t)\right] \psi (x,t)$. Hence, if we will restrict
ourselves to the first order differential operators we should not include
the derivative $\partial _t$ in $L$ and the general form of this operator
should be as follows: $L=a(x,t)+b(x,t)\partial _x$.

We refer the reader for details of calculations to Refs. \cite{BS} and \cite
{BScs} and will give only the final result. The transformation operator $L$
has the form 
\begin{equation}
\label{L}L=L_1(t)[-u_x(x,t)/u(x,t)+\partial _x],
\end{equation}
where the subscript $x$ denotes the partial derivative with respect to the
coordinate $x$ and 
\begin{equation}
\label{L1}L_1\left( t\right) =\exp \left[ 2\int \mathop{\rm d}t%
\mathop{\rm Im}\left( \log u\right) _{xx}\right] .
\end{equation}

The potential $V_1(x,t)$ is defined by the potential difference 
\begin{equation}
\label{A}A_{0,1}(x,t)=h_1-h_0=V_1(x,t)-V_0(x,t)=-[\log \left| u(x,t)\right|
^2]_{xx}. 
\end{equation}
Note that the operator $L$ and the new potential $V_1(x,t)$ are completely
defined by a function $u(x,t)$ called {\it the transformation function}.
This function is a particular solution of the initial Schr\"o\-din\-ger
equation (\ref{sc1}) subject to the condition

\begin{equation}
\label{rc}\left( \log \frac u{u^{*}}\right) _{xxx}=0,
\end{equation}
called {\it the reality condition of the new potential }(the ''*'' means
complex conjugation).

Operator $L^{+}$ which is Laplace adjoint to $L$%
\begin{equation}
\label{Lc}L^{+}=-L_1(t)[u_x^{*}(x,t)/u^{*}(x,t)+\partial _x]
\end{equation}
realizes the transformation in the inverse direction, i.e. the
transformation from the solutions of Eq. (\ref{sc2}) to the ones of Eq. (\ref
{sc1}). As a result, the product $L^{+}L$ is a symmetry operator for Eq. (%
\ref{sc1}) and $LL^{+}$ is the analogous one for Eq. (\ref{sc2}).

The action of the operator $L$ (\ref{L}) on the transformation function $u$
gives zero. However if this function is chosen in accordance with the Eq. (%
\ref{rc}) we can readily verify that the function 
\begin{equation}
\label{v}v(x,t)=\frac 1{L_1(t)u^{*}(x,t)}
\end{equation}
is the solution of the new Schr\"odinger Eq. (\ref{sc2}). Note the following
property of this function: $L^{+}v=0$.

With the help of the transformation operators $L$ and $L^{+}$ we build up
the time-dependent nilpotent supercharge operators 
\begin{equation}
\label{Q}Q=\left( 
\begin{array}{cc}
0 & 0 \\ 
L & 0
\end{array}
\right) ,\quad Q^{+}=\left( 
\begin{array}{cc}
0 & L^{+} \\ 
0 & 0
\end{array}
\right) 
\end{equation}
which commute with the Schr\"odinger super-operator $iI\partial _t-H$ where $%
H=\mathop{\rm diag}\{h_0,h_1\}$ is time-dependent super-Hamiltonian and $I$
is the unit $2\times 2$ matrix. In general, the super-Hamiltonian is not
integral of motion for the quantum system guided by the matrix
Schr\"o\-din\-ger equation 
\begin{equation}
\label{mse}\left( iI\partial _t-H\right) \Psi \left( x,t\right) =0.
\end{equation}
The two-component function $\Psi \left( x,t\right) $ belongs to the linear
space $T_{01}$ defined over complex number field and for every $\psi \in T_0$
spanned by the basis $\Psi _{+}=\psi e_{+}$, $\Psi _{-}=L\psi e_{-}$ with $%
e_{+}=\left( 1,0\right) ^T$ and $e_{-}=\left( 0,1\right) ^T$, the
superscript ''$T$'' stands for transposition.

The operators (\ref{Q}) are integrals of motion for Eq. (\ref{mse}). Using
the symmetry operators $L^{+}L$ and $LL^{+}$ we can construct another
integral of motion for this equation: $S=\mathop{\rm diag}\left\{
L^{+}L,LL^{+}\right\} $. The operators $Q$, $Q^{+}$ and $S$ realize the
well-known superalgebra \cite{Wit} $sl(1/1)$ but instead of Hamiltonians we
see in its construction another symmetry operators. In general, the operator 
$S$ (as well as $Q$) depends on time and consequently we have the
time-dependent superalgebra.

{\bf 3.} The majority of cases of physical interest are such that it is
possible to introduce the Hilbert space structure $H^0(R)$ in the space $T_0$
with the appropriately defined scalar product. We suppose that the operator $%
L^{+}$ is the operator Hermitian conjugate to $L$ with respect to this
scalar product. The symmetry operator $L^{+}L$ for $L$ being of the form (%
\ref{L}) is of the second degree in $\partial _x$. By means of Eq. (\ref{sc1}%
) the second derivative with respect to $x$ for any $\psi \in T_0$ can be
replaced by the first derivative with respect to $t$. This means that the
restriction of the operator $L^{+}L$ on the space $T_0$ is an operator of
the first degree in $\partial _x$ and $\partial _t$ and consequently the
operator $iL^{+}L$ belongs to the real Lie algebra $G_0$ of symmetry
operators for Eq. (\ref{sc1}) which we assume to be known. Since the
transformation function $u$ is an eigenfunction of the operator $iL^{+}L$
corresponding to the zero eigenvalue we can denote $L^{+}L=ig^{(0)}-\alpha _0
$ with $g^{(0)}\in G_0$ and $ig^{(0)}u=\alpha _0u,$ $\alpha _0\in {\Bbb R}$.
The operator $g^{(0)}$ is usually skew-Hermitian and $ig^{(0)}$ has a unique
self-adjoint extension (see, for example, \cite{Mil}). In $H^0(R)$ this
operator may have discrete spectra, continuous spectra or a combination of
the two. The same also apply for the operator $LL^{+}=ig^{(1)}-\alpha _0$, $%
g^{(1)}\in G_{1},$ $G_1$ being a Lie algebra of symmetry operators
for Eq. (\ref{sc2}). Operator $g^{(1)}$ is said to be {\it a superpartner}
of $g^{(0)}$. It is not difficult to calculate for the transformation
function $u$ subject to the reality condition (\ref{rc}) the commutator $%
L^{+}L-LL^{+}=L_1^2(t)[\log \left| u(x,t)\right|
^2]_{xx}=L_1^2(t)(h_0-h_1)=-L_1^2(t)A_{0,1}(x,t)$. So we have $%
g^{(0)}-g^{(1)}=iL_1^2(t)A_{0,1}(x,t)$.

If the transformation function $u$ is nodeless, the potential difference (%
\ref{A}) has no severe singularities and the operators $L$ (\ref{L}) and $%
L^{+}$ (\ref{Lc}) are well defined. There exists a sole (up to a constant
factor) nodeless function in the space $H^0(R)$. Beyond this space there are
many nodeless functions suitable for use as transformation function.

Let us assume that the nodeless transformation function $u$ is such that the
absolute value of $u^{-1}(x,t)$ is square integrable. Then the solution of
Eq. (\ref{sc2}) of the form (\ref{v}) is the discrete spectrum eigenfunction
of $g^{(1)}$. In the space $T_1$ the operator $L$ induces a subspace $%
H^{11}(R)=\{\varphi :$ $\varphi =L\psi ,$ $\forall \psi \in H^0(R)\}$. If we
denote $H^{10}(R)=\ker L^{+}=\{\gamma v,$ $\forall \gamma \in {\Bbb C}\}$
then we have the decomposition of the Hilbert space $H^1(R)=H^{10}(R)\oplus
H^{11}(R)\subset T_1$. In this case the lowest eigenvalue of the
super-operator $S$ is nondegenerate and its eigenfunction $\Psi _0=N\left(
0,v\right) ^T$ ($N$ being the normalization constant) represents the vacuum
state and is annihilated by both supercharges $Q$, $Q^{+}$ and,
consequently, we have the exact supersymmetry.

We can repeat all these reasonings starting from the Eq. (\ref{sc2}),
Hamiltonian $h_1$ and symmetry operator $g^{(1)}$ and construct new
Hamiltonian $h_2=h_1+A_{1,2}(x,t)$, symmetry operator $%
g^{(2)}=g^{(1)}-iL_{1,2}^2(t)A_{1,2}(x,t)$, and new exactly solvable
Schr\"o\-din\-ger equation etc. The chain of $N$ time-dependent Darboux
transformations can serve us as a basis for construction of $N$-extended
supersymmetric model or parasupersymmetric one. These constructions can be
made in full analogy with the corresponding stationary constructions \cite
{Pash}-\cite{pssqm}, \cite{TMF}.

4. Let we have a chain of the well defined Hamiltonians $h_0\rightarrow
h_1\rightarrow \ldots \rightarrow h_N$ obtained by the subsequent
applications Darboux transformation operators $L_{0,1}$, $L_{1,2}$, \ldots , 
$L_{N-1,N}$. With the help of Eq. (\ref{L}) we can express all the
intermediate transformation functions through the solutions of the initial
equation (\ref{sc1}). As a result, we eliminate all the intermediate
Hamiltonians and Schr\"o\-din\-ger equations and pass from Eq. (\ref{sc1})
immediately to the final one. The chain of the first-order transformations
collapses in a single $N$-order transformation and we obtain a nonstationary
analogue of the well known Crum-Krein formula \cite{Crum}, \cite{Krein}:%
$$
L_{0,N}=L_{N-1,N}L_{N-2,N-1}\ldots L_{0,1}=L_N(t)W(u_1,u_2,\ldots
,u_N)\left| 
\begin{array}{cccc}
u_1 & u_2 & \ldots  & 1 \\ 
u_{1x} & u_{2x} & \ldots  & \partial _x \\ 
\vdots  & \vdots  & \ddots  & \vdots  \\ 
u_{1x}^{(N)} & u_{2x}^{(N)} & \ldots  & \partial _x^{(N)}
\end{array}
\right|  
$$
where the conventional symbol $W(u_1,u_2,\ldots ,u_N)$ stands for the
Wronskian of the transformation functions $u_1,u_2,\ldots ,u_N$. If there is
no necessity to introduce the intermediate Hamiltonians and
Schr\"o\-din\-ger equations we can avoid the reality condition for every
transformation function $u_i$ and impose a single condition of reality of
the final potential 
\begin{equation}
\label{rc2}\left[ \log \left( \frac{W(u_1,u_2,\ldots ,u_N)}{%
W^{*}(u_1,u_2,\ldots ,u_N)}\right) \right] _{xxx}=0.
\end{equation}
The potential difference for this case reads as follows%
$$
A_{0,N}=-\left[ \log \left| W(u_1,u_2,\ldots ,u_N)\right| ^2\right] _{xx} 
$$
and function $L_N(t)$ can be chosen real%
$$
L_N(t)=\exp \left( 2\int dt\mathop{\rm Im}\left[ \log W(u_1,u_2,\ldots
,u_N)\right] _{xx}\right) . 
$$
We want to stress that the condition (\ref{rc}) may be violated but the
condition (\ref{rc2}) should hold. The analogous case for the stationary
Schr\"o\-din\-ger equation is called {\it the irreducible} one \cite{A} and
the operator $L_{0,N}$ is said to be {\it the }${\it N}${\it th-order
Darboux transformation operator} \cite{TMF}.

{\bf 5.} Consider the case when all the transformation functions $u_i$, $%
i=1,2,\ldots ,N$ are eigenfunctions of the same symmetry operator $g^{(0)},$ 
$ig^{(0)}u_k=\alpha _ku_k$, $\alpha _1>\alpha _2>\ldots >\alpha _N$ and
every Wronskian $W(u_1,u_2,\ldots ,u_p)$, $p=1,2,\ldots ,N$, $W(u_1)\equiv
u_1$ conserves its sign in {\em R} for all $t$ and satisfies the reality
condition (\ref{rc2}). Together with the chain of the Schr\"o\-din\-ger
equations with the well-defined Hamiltonians $h_i$ we now have the chain of
the well-defined symmetry operators $g^{(i)}$. We call this case {\it the
completely reducible }one. Each pair of the adjacent operators $g^{(p)}$ and 
$g^{(p+1)}$ as well as the Schr\"o\-din\-ger equations with the adjacent
Hamiltonians $h_p$ and $h_{p+1}$ is intertwined by the well-defined
first-order transformation operator $L_{p,p+1}$, $p=0,1,\ldots ,N-1$.
Consider the $n$th-order ($n=1,2,\ldots ,N$) operators $%
L_{p,p+n}=L_{p+n-1,p+n}\ldots L_{p+1,p+2}L_{p,p+1}$. It is evident that
every operator $L_{p,p+n}$ intertwines two Schr\"o\-din\-ger equations with
the Hamiltonians $h_p$, $h_{p+n}$ and two symmetry operators $g^{(p)}$, $%
g^{(p+n)}$ and consequently is the $n$th-order Darboux transformation
operator. If we denote the operator Laplace adjoint to $L_{p,p+n}$ by $%
L_{p,p+n}^{+}$ then we can establish the following factorization properties: 
\begin{equation}
\label{f}
\begin{array}{c}
L_{p,p+n}^{+}L_{p,p+n}^{}=\prod\limits_{k=1}^n\left( ig^{(p)}-\alpha
_{p+k}\right) , \\ 
L_{p,p+n}L_{p,p+n}^{+}=\prod\limits_{k=1}^n\left( ig^{(p+n)}-\alpha
_{p+k}\right) .
\end{array}
\end{equation}

The whole chain of the transformations is associated with supercharges $%
Q_{p,q}=L_{p,q}e_{p,q}$, $p<q$; $p,q=0,1,\ldots ,N$, super-Hamiltonian $H=%
{\rm \mathop{\rm diag}}\{h_0,h_1,\ldots ,h_N\}$ and super-operator $S={\rm 
\mathop{\rm diag}}\{ig^{(0)},ig^{(1)},\ldots ,$ $ig^{(N)}\}$. The notations $%
e_{p,q}$ (see \cite{BD}) refer to $\left( N+1\right) $-dimensional square
matrix whose rows and columns are labeled from $0$ to $N$ and contain zeros
everywhere except units at the intersection of the $p$th column and $q$th
row. As $Q_{p,q}^{+}$ we denote the adjoint operators $%
Q_{p,q}^{+}=L_{p,q}^{+}e_{q,p}$. The chain of the $N+1$ Schr\"o\-din\-ger
equations is summarized in a single matrix equation (\ref{mse}) where $I$ is
the $\left( N+1\right) $-dimensional unit matrix. The intertwining relations
between the operators $L_{p,p+n}$, $L_{p,p+n}^{+}$ and the Schr\"o\-din\-ger
operators $i\partial _t-h_p$, $i\partial _t-h_{p+n}$ translate into the
commutation relations between $Q_{p,p+n}$ and matrix Schr\"o\-din\-ger
operator $iI\partial _t-H$. Hence, the supercharges $Q_{p,q}$ and $%
Q_{p,q}^{+}$ are integrals of motion for the system with the
super-Hamiltonian $H$ and the symmetry operator $S$ commutes with all $%
Q_{p,q}$'s and $Q_{p,q}^{+}$'s. The reducibility condition leads to the
nonlinear algebra of the following relations: 
\begin{equation}
\label{psa}
\begin{array}{c}
Q_{p,q}Q_{s,p}=Q_{s,q},\quad s<p<q,\quad s,p,q=0,1,\ldots N, \\ 
Q_{p,p+n}Q_{p,p+n+m}^{+}=\prod\limits_{i=1}^n(S-\alpha
_{p+i})Q_{p+n,p+n+m}^{+},\quad p+n+m\leq N, \\ 
Q_{p-n-m,p}^{+}Q_{p-n,p}=\prod\limits_{i=1}^n(S-\alpha
_{p+i-1})Q_{p-n-m,p-n}^{+},\quad p-n-m\geq 0,
\quad p\leq N, \\ Q_{p,p+n}Q_{p,p+n}^{+}Q_{p,p+n}=\prod%
\limits_{i=1}^n(S-\alpha _{p+i})Q_{p,p+n}
,\quad p+n\leq N, \\ n,m=1,2,\ldots 
\end{array}
\end{equation}
and of the Hermitian conjugated ones. All the other products of every two
supercharges are equal to zero. Hence we obtain the nonlinear algebra with a
parasupersymmetric structure.

{\bf 6.} As an example consider the free particle parasupersymmetric model: $%
V_0(x,t)=0,$ $R={\Bbb R}$. Functions $\psi _n(x,t)$ of the discrete basis
set in the space $H^0(R)$ are eigenfunctions of the operator \cite{Mil}: $%
g^{(0)}=K_{-2}-K_2=i(1+t^2)\partial _x^2+tx\partial _x+t/2-ix^2/4$, $%
g^{(0)}\psi _\lambda =i\lambda \psi _\lambda $ and correspond to $\lambda
=\lambda _n=-n-1/2$, $n=0,1,2,\ldots $ ($\psi _n\equiv \psi _{\lambda _n}$).
Their coordinate representation is as follows \cite{Mil}: 
\begin{equation}
\label{ff}
\begin{array}{c}
\psi _\lambda (x,t)=(1+t^2)^{-1/4}\exp [ix^2t/(4+4t^2)+i\lambda \arctan
t]Q_\lambda (z)
, \\ z=x/\sqrt{1+t^2}
\end{array}
\end{equation}
where $Q_\lambda (z)$ is the parabolic cylinder function satisfying the
equation 
$$
Q_\lambda ^{\prime \prime }(z)-(z^2/4+\lambda )Q_\lambda (z)=0, 
$$
where prime denotes the derivative with respect to $z$. It is easy to see
that the functions (\ref{ff}) for any real $\lambda $ satisfy the reality
condition (\ref{rc}).

Functions (\ref{ff}) at $\lambda =\lambda _m=m+1/2$, $m=0,1,2,\ldots $ do
not belong to the space $H^0(R)$ and at even $m$ are nodeless for all $t\in 
{\Bbb R}$ and suitable for use for the single Darboux transformation as
transformation functions $u_m=\psi _{\lambda _m}$. (Note that for odd $m$
they are equal to zero at $x=0$). Using formulas (\ref{L}-\ref{L1}) we find
the transformation operator%
$$
L=L_{0,1}=L_1(t)\partial _x-\frac x2\sqrt{\frac{1+it}{1-it}}-im\frac{%
He_{m-1}(iz)}{He_m(iz)},\quad L_1(t)=\sqrt{1+t^2} 
$$
where $He_m(z)=2^{-m/2}H_m(z/\sqrt{2})$ ($H_m(z)$ are Hermite polynomials).
New Schr\"o\-din\-ger equation potential is calculated by the Eq. (\ref{A})%
$$
V_1=A_{0,1}=(1+t^2)^{-1}[2m(m-1)\frac{He_{m-2}(iz)}{He_m(iz)}-2m^2\frac{%
He_{m-1}^2(iz)}{He_m^2(iz)}-1]. 
$$
The same potential difference defines the symmetry operator $%
g^{(1)}=g^{(0)}-iL_1^2(t)A_{0,1}$ of the Schr\"o\-din\-ger equation with the
potential $V_1$. Its discrete eigenfunctions obtained with the help of the
operator $L_{0,1}$ and with the use of formula (\ref{L1}) form the basis of
the space $H^1(R)$ of the square integrable solutions of the new
Schr\"o\-din\-ger equation 
$$
\begin{array}{c}
\varphi _0=
\sqrt{m!}(2\pi )^{-1/4}[L_1(t)u_m^{*}(x,t)]^{-1}, \\ g^{(1)}\varphi
_0=i(m+1/2)\varphi _0, \\ 
\varphi _{n+1}=[n!(n+m+1)(1+it)
\sqrt{2\pi }]^{-1/2}\exp [-x^2/(4+4it)-in\arctan t] \\ \times
[He_{n+1}(z)+imHe_n(z)
{{\frac{He_{m-1}(iz)}{He_m(iz)}}}], \\ g^{(1)}\varphi
_{n+1}=-i(n+1/2)\varphi _{n+1},\quad n=0,1,2,\ldots .
\end{array}
$$
The normalisation constants of the functions $L_{0,1}\psi _n$, $n=0,1,\ldots 
$ are found with the help of the following relatrions:%
$$
\begin{array}{c}
\langle L_{0,1}\psi _n\mid L_{0,1}\psi _n\rangle =\langle \psi _n\mid
L_{0,1}^{+}L_{0,1}\mid \psi _n\rangle  \\ 
=\langle \psi _n\mid ig^{(0)}+m+1/2\mid \psi _n\rangle =(n+m+1)\langle \psi
_n\mid \psi _n\rangle .
\end{array}
$$
The normalisation constant for the function $\varphi _0$ is obtained by the
direct calculation.

It can be shown that the action of the operator $L_{0,1}$ on the functions (%
\ref{ff}) with $\lambda =\lambda _l=l+1/2$, $l=m+1,$ $m+3,\ldots $ gives
nodeless solutions of the new Schr\"o\-din\-ger equation 
$$
\begin{array}{c}
v_l(x,t)=L_{0,1}\varphi _{\lambda _l}(x,t)=(1-it)^{-1/2}\exp
[x^2/(4-4it)+il\arctan t]
{{\frac{f_{m,l}(z)}{He_m(iz)}}}, \\ 
f_{m,l}(z)=i[He_l(iz)He_{m+1}(iz)-He_m(iz)He_{l+1}(iz)]
\end{array}
$$
which do not belong to the space $H^1(R)$ and are suitable for use for the
second transformation as the transformation functions. Note that the
absolute value of the function $v_l^{-1}(x,t)$ is square integrable for all $%
t$ and $l=m+1,m+3,\ldots $. This fact can easily be shown by the direct
calculation of their normalisation constants. The transformation operator $%
L_{1,2}$ constructed with the help of the transformation function $%
u(x,t)=v_l(x,t)$ according with the formulas (\ref{L} - \ref{L1}) has the
form%
$$
L_{1,2}=\sqrt{1+t^2}\partial _x-\frac x2\sqrt{\frac{1+it}{1-it}}+im\frac{%
He_{m-1}(iz)}{He_m(iz)}-i\frac{f_{ml}^{\prime }(z)}{f_{ml}(z)}. 
$$
New potential difference $A=A_{1,2}=V_2-V_1$ is anew calculated by the Eq. (%
\ref{A}) and leads to a new exactly solvable potential%
$$
V_2=-2(1+t^2)^{-1}[1+\frac{f_{ml}^{\prime \prime }(z)}{f_{ml}(z)}-\left( 
\frac{f_{ml}^{\prime }(z)}{f_{ml}(z)}\right) ^2]. 
$$
New Schr\"odinger equation symmetry operator has the form $%
g^{(2)}=g^{(1)}-iL_1^2(t)A_{1,2}=g^{(0)}-iL_1^2(t)A_{0,2}=g^{(0)}-iL_1^2(t)V_2
$. (Note that since $u_l/u_l^{*}=v_l/v_l^{*}$, the time-dependent factor $%
L_1(t)$ calculated by the formula (\ref{L1}) is always the same for this
case). Its orthonormal eigenfunctions $\chi _{i+1}(x,t)$, $i=0,1,2,\ldots $
forming together with the function $\chi _0(x,t)\sim \left[
L_1(t)v_l^{*}(x,t)\right] ^{-1}$ the discrete basis set of the space $H^2(R)$
of the square integrable solutions of the Schr\"odinger equation with the
potential $V_2$ should be found by applying the operator $L_{1,2}$ on the
functions $\varphi _i(x,t)$. Finally we obtain 
$$
\begin{array}{c}
\chi _0=[(2\pi )^{-1/2}l!(l-m)(1-it)^{-1}]^{1/2} \\ 
\times \exp [-x^2/(4+4it)+il\arctan t]
{\displaystyle {\frac{He_m(iz)}{f_{ml}(z)}}}, \\ \chi _1=[(2\pi
)^{-1/2}m!(l-m)(1-it)^{-1}]^{1/2} \\ 
\times \exp [-x^2/(4+4it)+im\arctan t]
{\displaystyle {\frac{He_l(iz)}{f_{ml}(z)}}}, \\ \chi
_{n+2}=[(n+l+1)(n+m+1)]^{-1/2} \\ 
\times [-(l+n+1)\varphi _n+(l-m)u_m
{\displaystyle {\frac{W(\varphi _n,u_l)}{W(u_m,u_l)}}}], \\ n=0,1,2,\ldots ,
\end{array}
$$
where%
$$
\begin{array}{c}
W(\varphi _n,u_l)=(n!
\sqrt{2\pi })^{-1/2}(1+t^2)^{-1}\exp [itx^2/(2+2t^2)+i(l-n)\arctan t]\times 
\\ \lbrack nHe_{n-1}(z)He_l(iz)+iHe_n(z)He_{l+1}(iz)]
\end{array}
$$
and%
$$
W(u_m,u_l)=\left( (1-it)\sqrt{1+t^2}\right) ^{-1}f_{ml}(z)\exp
[x^2/(2-2it)+i(m+l)\arctan t]. 
$$

Second-order operator $L_{0,2}=L_{1,2}L_{0,1}$ intertwines the initial
Schr\"o\-din\-ger equation (with the zero potential) and the final one (with
the potential $V_2$) and participates in the following factorizations [see
Eq. (\ref{f})]: $L_{0,2}^{+}L_{0,2}=(ig^{(0)}+m+1/2)(ig^{(0)}+l+1/2)$ and $%
L_{0,2}L_{0,2}^{+}=(ig^{(2)}+m+1/2)(ig^{(2)}+l+1/2)$. The operators $L_{0,1}$%
, $L_{1,2}$ and $L_{0,2}$ form the complete set of the transformation
operators for the case $N=2$. We associate with this set the complete set of
the supercharges $Q_{p,q}=L_{p,q}e_{p,q}$ and $Q_{p,q}^{+}=(Q_{p,q})^{%
\dagger }$, $p,q=0,1,2$ ($p<q$), which together with the super-operator $S=i%
\mathop{\rm diag}\{g^{(0)},g^{(1)},g^{(2)}\}$ form the following nonlinear
algebra [see formulae (\ref{psa})]: 
$$
\begin{array}{c}
\lbrack S,Q_{p,q}]=0,\quad Q_{1,2}Q_{0,1}=Q_{0,2}, \\ 
Q_{0,1}Q_{0,1}^{+}Q_{0,1}=(S+m+1/2)Q_{0,1}, \\ 
Q_{1,2}Q_{1,2}^{+}Q_{1,2}=(S+l+1/2)Q_{1,2}, \\ 
Q_{0,2}Q_{0,2}^{+}Q_{0,2}=(S+m+1/2)(S+l+1/2)Q_{0,2}, \\ 
Q_{0,1}Q_{0,2}^{+}=(S+m+1/2)Q_{1,2}^{+}, \\ 
Q_{0,2}^{+}Q_{1,2}=(S+l+1/2)Q_{0,1}^{+},
\end{array}
$$
and relations Hermitian conjugated to these. All the other products of any
two supercharges are equal to zero.

The lowest eigenvalue $-l-1/2$ of the superoperator $S$ is nondegenerate.
Its eigenfunction (vacuum state) has the form $\Psi _0=(0,0,\chi _0)^T$. Two
lowest excited states of this operator [$\Psi _1^1=(0,0,\chi _1)^T$, $\Psi
_1^2=(0,\varphi _0,0)^T$] correspond to the eigenvalue $-m-1/2$. All the
other eigenvalues are triple degenerate and equal $n+1/2$, $n=0,1,2,\ldots $%
. Their eigenfunctions have the form $\Psi _{n+2}^1=\left( 0,0,\chi
_{n+2}\right) ^T$, $\Psi _{n+2}^2=\left( 0,\varphi _{n+1},0\right) ^T$, $%
\Psi _{n+2}^3=\left( \psi _n,0,0\right) ^T$.

\end{document}